\documentclass[prb,preprint,superscriptaddress]{revtex4-1} 

\usepackage{amsmath}  
\usepackage{amsfonts} 
\usepackage{graphicx} 
\usepackage{esvect}
\usepackage{mathtools}
\usepackage{subcaption}
\usepackage{xcolor}
\usepackage{appendix}

\usepackage{siunitx} 
\usepackage{placeins}

\usepackage{float}

\usepackage{pbox}

\usepackage{url}

\usepackage{xcolor}
\pagecolor{white}

\begin{document}


\title{Infinite AC Ladder with a ``Twist''}

\author{Quan M. Nguyen}
\affiliation{Hanoi-Amsterdam High School, 01 Hoang Minh Giam Str., Trung Hoa Nhan Chinh, Cau Giay Dist., Hanoi 100000, Vietnam.}

\author{Linh K. Nguyen}
\affiliation{Lam Son High School, 307 Le Lai Str., Dong Son Dist., Thanh Hoa 440000, Vietnam}
\affiliation{Massachusetts Institute of Technology, Cambridge, MA 02139.}

\author{Tung X. Tran}
\affiliation{Hanoi-Amsterdam High School, 01 Hoang Minh Giam Str., Trung Hoa Nhan Chinh, Cau Giay Dist., Hanoi 100000, Vietnam.}
\affiliation{Massachusetts Institute of Technology, Cambridge, MA 02139.}

\author{Chinh D. Tran}
\affiliation{Hung Vuong High School, 70 Han Thuyen Str., Tan Dan, Viet Tri, Phu Tho 35000, Vietnam.}

\author{Truong H. Cai}
\affiliation{School of Engineering, Brown University, Providence, RI 02912, USA.}

\author{Trung V. Phan}
\email{tvphan@princeton.edu}
\affiliation{Department of Physics, Princeton University, Princeton, NJ 08544, USA.}

\date{\today}

\begin{abstract}
The infinite AC ladder network can exhibit unexpected behavior. Entangling the topology brings even more surprises, found by direct numerical investigation. We consider a simple modification of the ladder topology and explain the numerical result for the complex impedance, using linear algebra. The infinity limit of the network's size corresponds to keeping only the eigenvectors of the transmission matrix with the largest eigenvalues, which can be viewed as the most dominant modes of electrical information that propagate through the network.
\end{abstract}

\maketitle 

\section{A Curious Finding from an Infinite AC Ladder with a ``Twist''}

\begin{figure*}[!htb]
\centering
\includegraphics[width=0.6\textwidth]{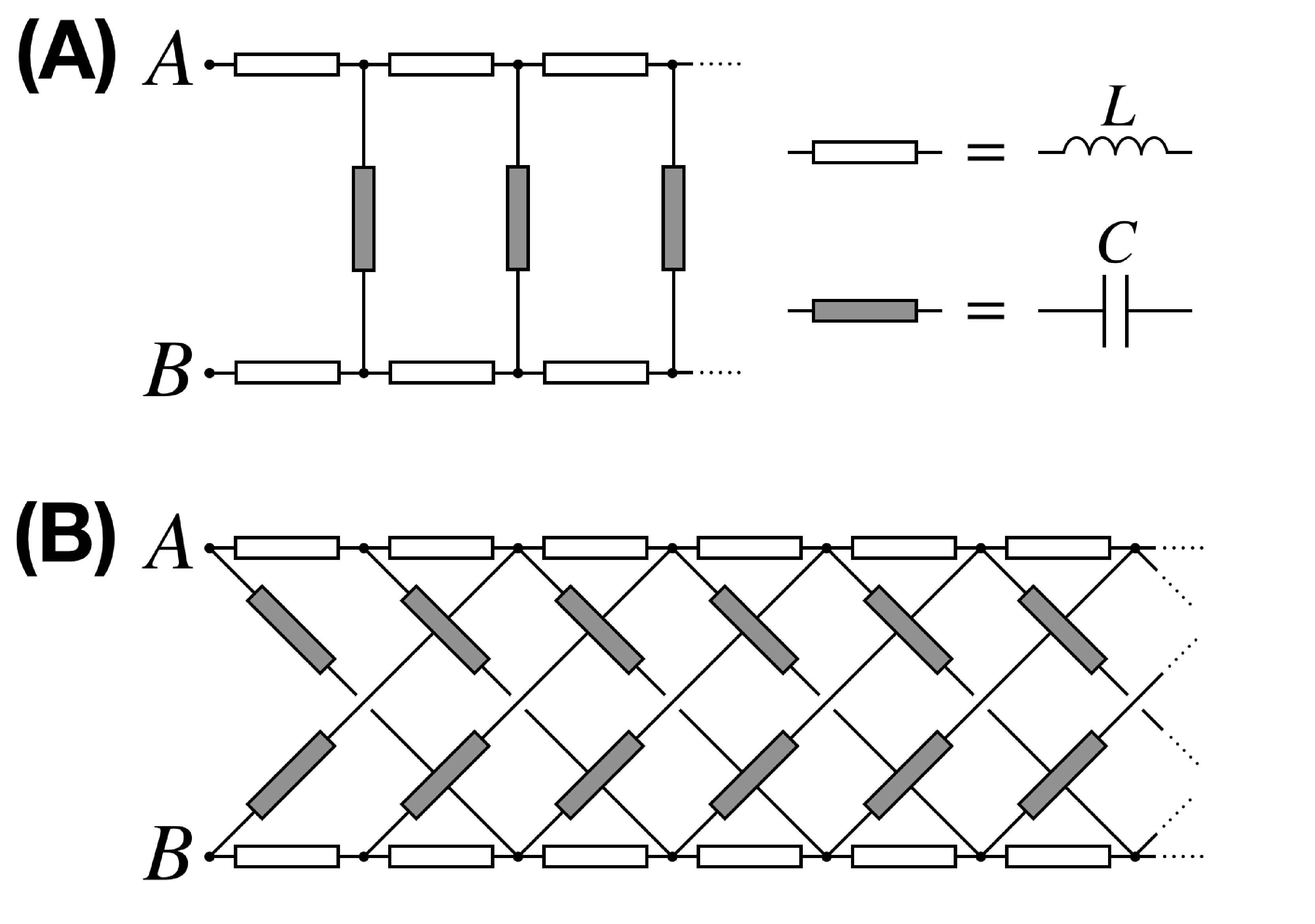}
\caption{(A) The infinite AC ladder network. (B) An infinite AC ladder network with a ``twist''. We call this topology the symmetric twisted ladder.}
\label{ladder_topology_ac}
\end{figure*}

The infinite AC network with ladder topology given in Fig. \ref{ladder_topology_ac}A is well-known, appears in many introductory physics courses and standard textbooks as a model of a transmission line \cite{feynman_lecture, ryder_network,jordan_em, ramo_fw}. It was even introduced to high school students in 18th International  Physics Olympiad (East Germany 1987). For the network consisting of identical inductors $L$ and capacitors $C$, by adding one more unit cell and assuming convergence, we obtain a consistency equation \cite{feynman_lecture}: \begin{equation}
Z_{AB}(\omega) = 2i\omega L + \frac{Z_{AB}(\omega)/i\omega C}{Z_{AB}(\omega) + 1/i\omega C} \ \ , \ \ 
\end{equation}
in which the complex impedance can be solved:
\begin{equation}
Z_{AB}(\omega) = \omega L \big( i + \sqrt{-1+ 2/\omega^2LC}\big) \ \ , \ \
\end{equation}
where $\omega$ is the AC frequency. This result exhibits two distinct behaviors in different ranges of $\omega$: (i) when $\omega > \sqrt{2/LC}$, $Z_{AB}(\omega)$ is purely imaginary; (ii) when $\omega \leq \sqrt{2/LC}$, $Z_{AB}(\omega)$ has both non-zero imaginary and real components, which is strange since all elements in the network have imaginary impedances. In fact, the consistency equation requires $Z_{AB}(\omega)$ to converge as the size of the networks goes to infinity $N\rightarrow \infty$, which is wrong in this range of frequency. However, by adding a very small resistance to every element in this network, $\text{Re}\big(Z_{AB}(\omega)\big)$ emerges and convergence reappears, which is shown elegantly in a numerical investigation by Van Enk et al \cite{Enk_LC} and later studied in greater theoretical detail \cite{yoon_ladder, ucak_convergence, ucak_understanding, Dykhne_stability}. 

\begin{figure*}[!htb]
\centering
\includegraphics[width=0.9\textwidth]{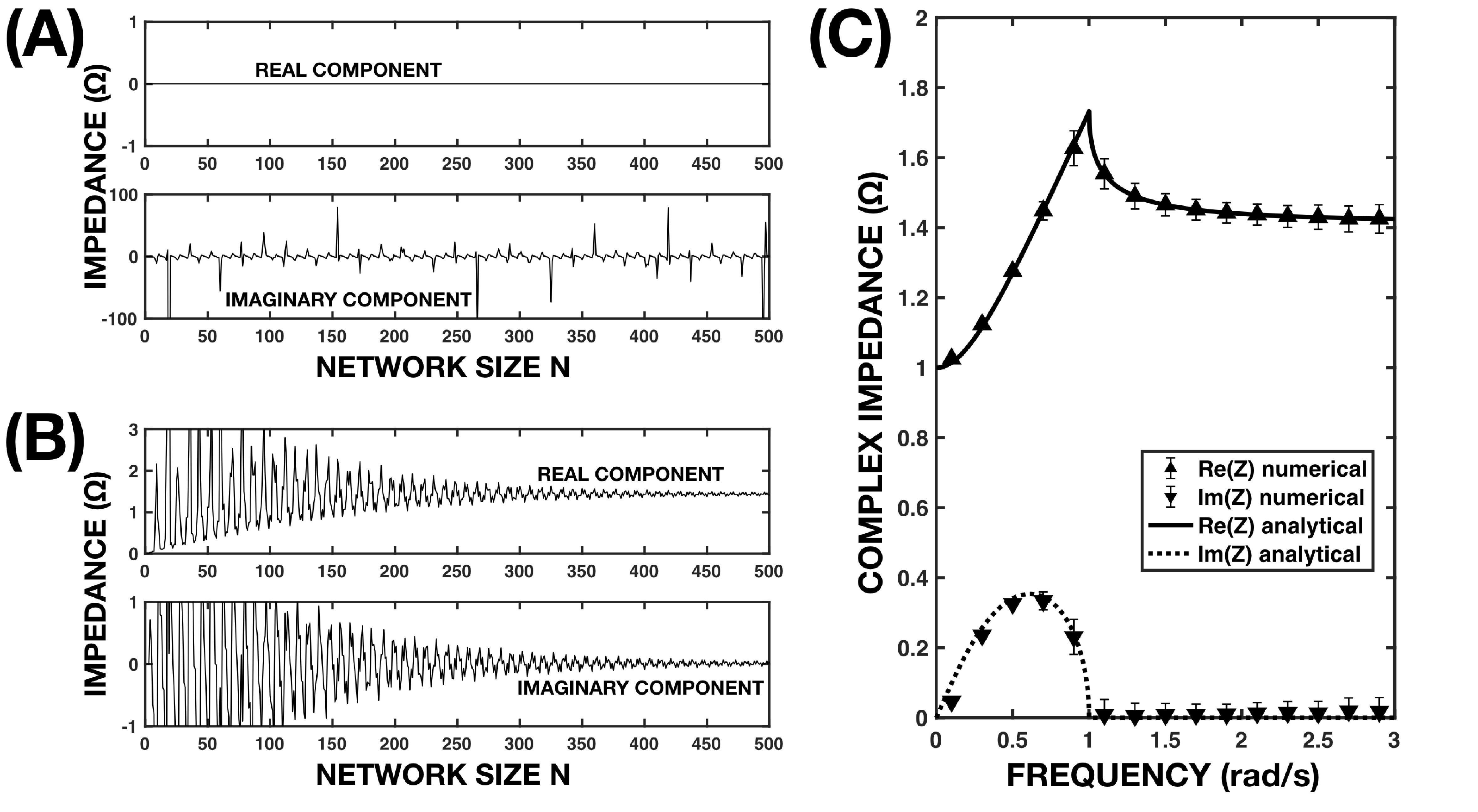}
\caption{A numerical investigation on the symmetric twisted ladder consists of inductors $L=1$H and capacitors $C=1$F. (A) The complex impedance versus the network size $N$, for frequency $\omega = 2$rad/s and all AC elements have no resistance. (B) The complex impedance versus the network size $N$, for frequency $\omega = 2$rad/s and all AC elements have resistance $R=0.01\Omega$. (C) The average complex impedance in the range $N\in[450,500]$ versus the network frequency, when all AC elements have resistance $R=0.01\Omega$. We also compare those with the analytical result for $N\rightarrow \infty$ and $R\rightarrow 0$ given in equation \eqref{theoretical_result}.}
\label{numerical_investigation}
\end{figure*}

In this paper, we explore how entangling the topology can change the behavior of the infinite AC network. Previous research was done on infinite lattice topologies \cite{tzeng_grid} and fractal topologies \cite{clerc_fractal, chen_fractal, alonso_fractal}. Here we consider a much simpler topology with as much complexity if not more: a modification of the ladder topology by ``adding a twist'', which we call a symmetric twisted ladder topology (see Fig. \ref{ladder_topology_ac}B). Understanding more about the properties of the ladder-like AC networks are of relevant to not only 
electrical engineering (eg. transmission line designs) but also biophysics, for example as a cochlear model \cite{ambi_digital} and a model of neural ionic-channels \cite{ghaffari_neuron}. We have found many surprises: not only is there no frequency range where $Z_{AB}(\omega)$ converges, but also adding infinitesimal resistances to all AC elements can make $Z_{AB}(\omega)$ converges to a positive real value, completely eliminating the imaginary component:
\begin{equation}
Z_{AB}(\omega)  \Big|_{\omega \geq 1/\sqrt{LC}} \in \mathbb{R}^+
\ \ , \ \ \lim_{\omega \rightarrow \infty} Z_{AB}(\omega) = \sqrt{2L/C} \ \ . \ \
\end{equation}
Our numerical finding is shown in Fig. \ref{numerical_investigation}. A theoretical explanation in detail for this curious behavior will be provided Section \ref{Section_Explanation}. 

\section{An Explanation with the Transmission Matrix Method \label{Section_Explanation}}

For linear-linking resistive networks, the voltages and currents of nodes in consecutive unit cells can be related by a transmission matrix \cite{Matthaei_Network}. We can apply this linear algebra method to find $Z_{AB}(\omega)$ of the symmetric twisted ladder.

\begin{figure*}[!htb]
\centering
\includegraphics[width=0.6\textwidth]{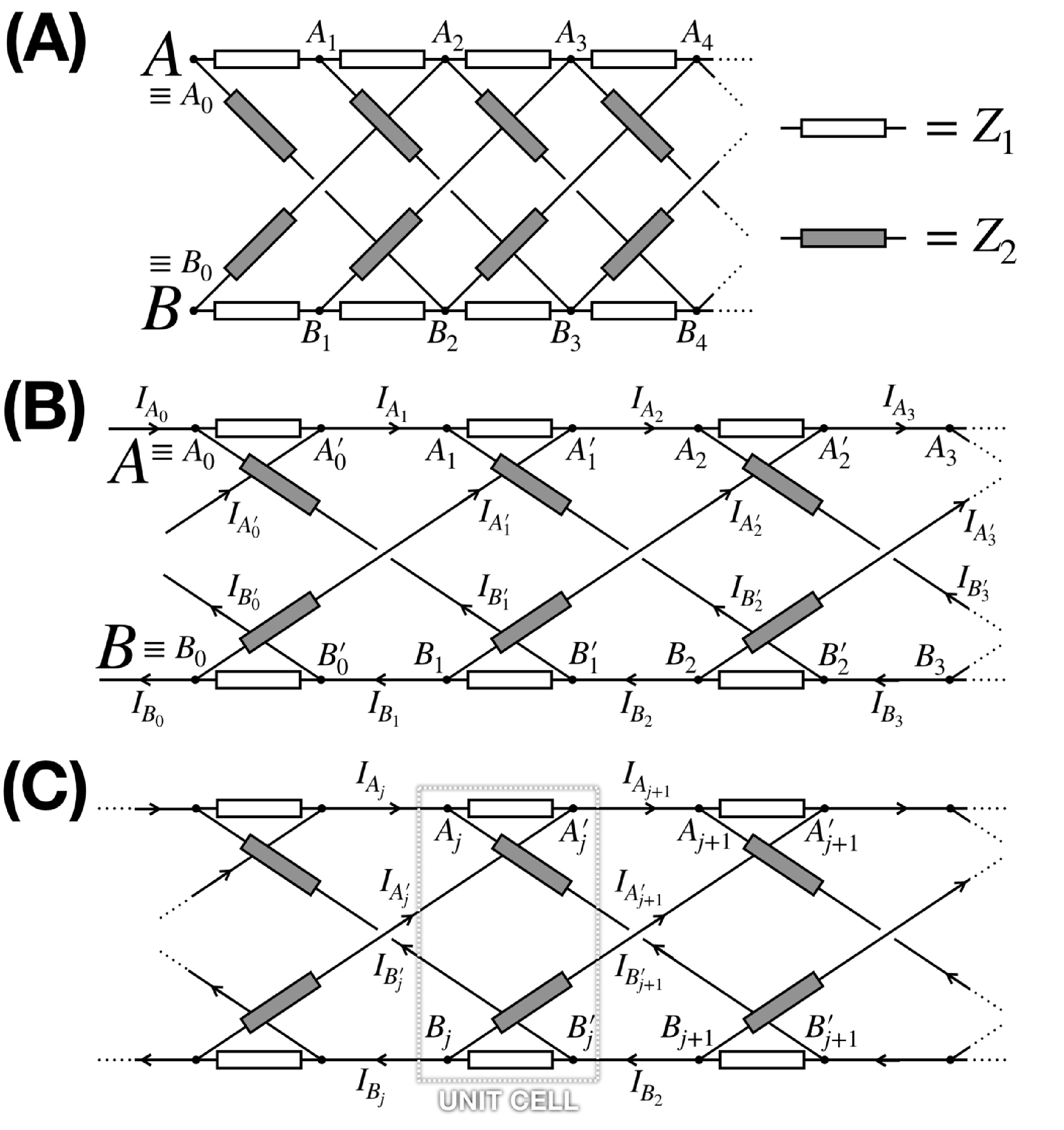}
\caption{(A) We label the nodes with $A_0$, $A_1$, $A_2$, ... and $B_0$, $B_1$, $B_2$, ... in which $A\equiv A_0$ and $B \equiv B_0$. (B) We split every points $A_j$, $B_j$ into $A_j$ and $A'_{j-1}$, $B_j$ and $B'_{j-1}$ connected by a wire of no resistance. Then we label the currents in those wires with $I_{A_j}$, $I_{B_j}$ and the currents flow in the diagonal resistors with $I_{A_j'}$ and $I_{B_j'}$. (C) An unit cell is a block of $A_jA'_j B'_j B_j$.}
\label{notation_dense_02}
\end{figure*}

We label the nodes as shown in Fig \ref{notation_dense_02}A. We split every point into two, then label the currents as shown in Fig \ref{notation_dense_02}B. The unit cells are now separated into blocks of $A_jA'_j B'_j B_j$ as shown in Fig \ref{notation_dense_02}C. The inlet current is $I_{A_0}$ and the outlet current is $I_{B_0}$, $I_{A_0}=I_{B_0}$. Due to symmetry, we have the currents $I_{A_j} = I_{B_j}$, $I_{A'_j} = I_{B'_j}$ and the voltages $V_{A_j}=-V_{B_j}$, $V_{A_j'}=-V_{B_j'}$. If an unit amplitude current comes into node $A$ and goes out from node $B$, then $I_{A_0} = I_{B_0} = 1$, $I_{A_0'} = I_{B_0'} = 0$ and $V_{A_0} = Z_{AB}(\omega)/2$.

\begin{figure*}[!htb]
\centering
\includegraphics[width=0.8\textwidth]{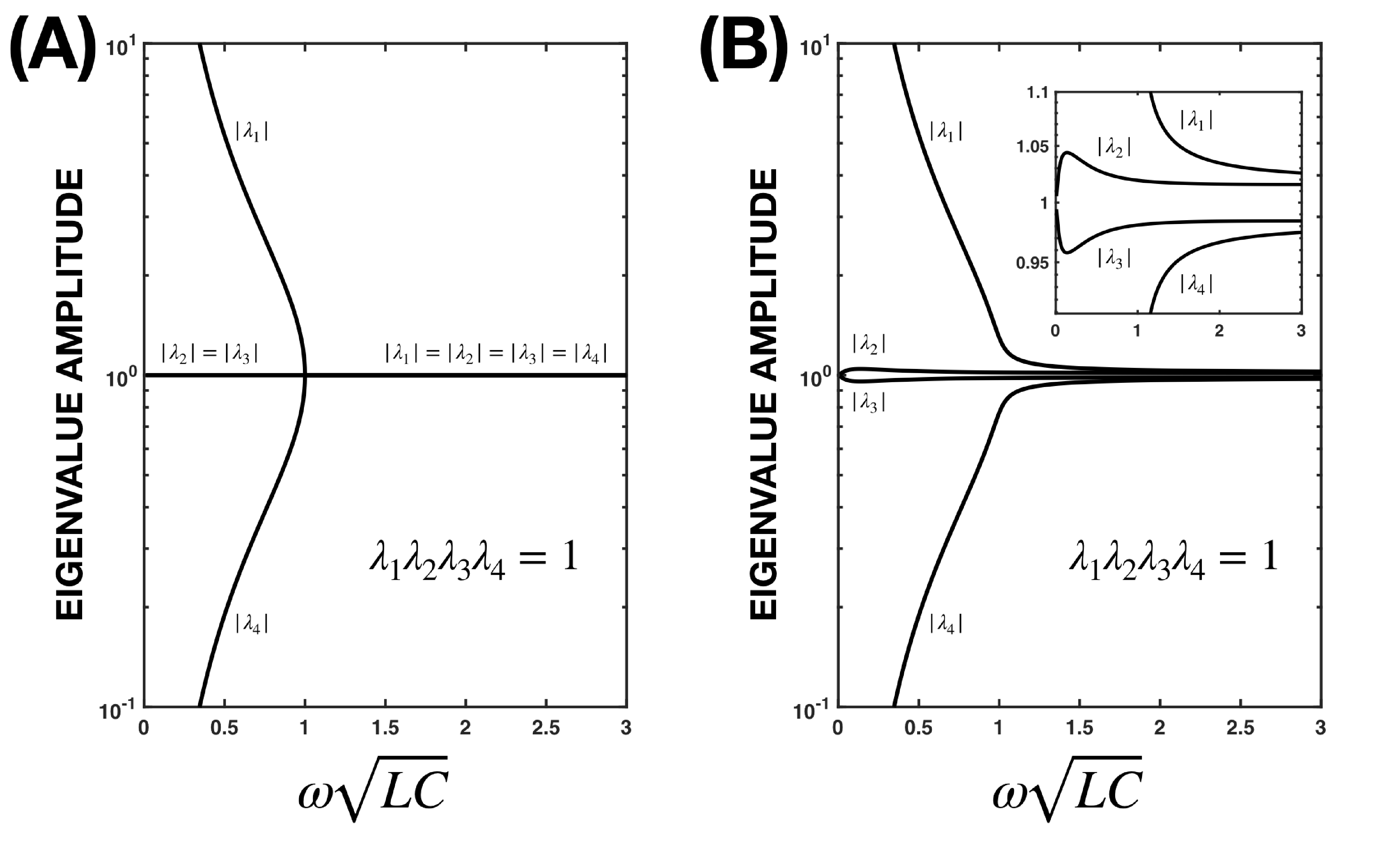}
\caption{The semilog plots represent how eigenvalue amplitudes of transmission matrix $[T]$ for the symmetric twisted ladder AC network depend on frequency $\omega$. (A) No resistance presented in the network. (B) A small resistance is added in series to every circuit elements.}
\label{eigenvalues}
\end{figure*}

The electrical information [E] transfer between two consecutive unit cells can be represented by a $4\times 4$ transmission matrix $[T]$:
\begin{equation}
[E_j] = [T] [E_{j+1}] \ \ , \ \ [E_j] = \begin{bmatrix}
     V_{A_j} \\ Z_1 I_{A_j} \\ V_{A'_j} \\ Z_1 I_{A'_j}
   \end{bmatrix} \ \ , \ \
\end{equation}
The transmission matrix can be found by applying Ohm's law and Kirchoff's laws to a unit cell, then solving for the components of $[E_j]$. Denoting $z = -Z_2/Z_1$, we obtained:
\begin{equation}
 [T] = \begin{bmatrix}
     0 & 0 & -1& z \\
     -1 & 0 & -1 & z-1 \\
     1 & 0 & 0 & 0 \\
     1 & 1 & 1 & -z
   \end{bmatrix} \ \ . \ \ 
\end{equation}
The transmission matrix $[T]$'s eigenvalues $\lambda$ and eigenvectors $[t]$ satisfy:
\begin{equation}
[T][t] = \lambda[t] \ \  \Rightarrow \ \ \lambda^4 + z\lambda^3 + 2(1-z)\lambda^2 + z\lambda + 1 = 0 \ \ . \ \ 
\end{equation} 
There are four complex solutions $\lambda_1$, $\lambda_2$, $\lambda_3$, $\lambda_4$ to this quartic polynomial, corresponding to four eigenvectors $[t_1]$, $[t_2]$, $[t_3]$, $[t_4]$. We will not write their expressions explicitly here, because not only are they very long and complicated but also we do not need their analytical forms to carry on the analysis. Without loss of generality, we assume that
\begin{equation}
|\lambda_1| \geq |\lambda_2| \geq |\lambda_3| \geq |\lambda_4| \ \ . \ \ 
\end{equation}
The components of the eigenvectors are in the same order of magnitude. From the symmetric properties of the polynomial, it can be deduced that $\lambda_1 \lambda_4=1$ and $\lambda_2 \lambda_3=1$. For a pure LC network with $Z_1 = i\omega L$ and $Z_2=1/i\omega C$, $\lambda_2 = \lambda_3^*$ and $[t_2]=[t_3]^*$. When $\omega \geq 1/\sqrt{LC}$ we even have all eigenvalue amplitudes to be equal, $\lambda_1=\lambda_4^*$ and $[t_1]=[t_4]^*$. If we add infinitesimally small resistance $0^+\Omega$ to every circuit elements then $z$ obtains a positive imaginary part $\text{Im(z)}=0^+$ and clear splitting between the eigenvalue amplitudes appears as shown in Fig. \ref{eigenvalues}:
\begin{equation}
|\lambda_1| > |\lambda_2| > 1 > |\lambda_3| > |\lambda_4| \ \ . \ \ 
\end{equation}

For a network of finite size $N$ there should be no current coming in from the end:
\begin{equation}
 [E_N] = \begin{bmatrix}
     V_{A_N} \\ 0 \\ V_{A'_N} \\ 0
   \end{bmatrix} \ \ , \ \ [E_0] = \begin{bmatrix}
     Z_{AB}^{(N)}/2\\ Z_1 \\ V_{A'_0} \\ 0 
   \end{bmatrix} \ \ . \ \ 
\end{equation}
Given that the eigenvalues are distinct, there is always a unique way to decompose $[E_N]$ into electrical information modes $[t]$ \cite{Strang_Linear_Algebra}:
\begin{equation} 
[E_N] = \sum^4_{k=1} c_k [t_k] \ \ , \ \ 
\end{equation}
where $c_1$, $c_2$, $c_3$, $c_4$ are complex numbers. Thus,
\begin{equation} 
[E_0] = [T]^N [E_0] = [T]^N[E_N] = \sum^4_{k=1} c_k \lambda_k^N [t_k] \ \ . \ \   
\end{equation} 
Solving the boundary conditions
\begin{equation}
\begin{split}
&\sum^4_{k=1} c_k [t_k]_2 = 0 \ \ , \ \ \sum^4_{k=1} c_k \lambda_k^N [t_k]_2 = Z_1 \ \ , \ \ 
\\
&\sum^4_{k=1} c_k [t_k]_4 = \sum^4_{k=1} c_k \lambda_k^N [t_k]_4 = 0  \ \ , \ \ 
\end{split}
\label{boundary_condition}
\end{equation},  
where $[t_k]_i$ denotes the i-th component of the k-th eigenvector, gives us the coefficients $c_1$, $c_2$, $c_3$, $c_4$, which can then be used to obtain
\begin{equation} 
Z_{AB}^{(N)} = 2Z_1 \frac{\sum^4_{k=1} c_k \lambda_k^N [t_k]_1 }{\sum^4_{k=1} c_k \lambda_k^N [t_k]_2}   \ \ . \ \ 
\end{equation}
The infinite network's complex impedance is defined from taking the limit of $N\rightarrow \infty$:
\begin{equation} 
Z_{AB} = \lim_{N\rightarrow \infty} Z_{AB}^{(N)} \ \ , \ \ 
\end{equation}
which only makes sense when $Z_{AB}^{(N)}$ is convergent.

\subsection{On the Emergence of Convergence}

To understand how convergence emerges, we start with a set of four eigenvalues with distinct amplitudes
\begin{equation} 
|\lambda_1| > |\lambda_2| > |\lambda_3| > |\lambda_4| \ \ , \ \
\end{equation}
At large $N\rightarrow \infty$ limit, the solution to the boundary conditions \eqref{boundary_condition} satisfies
$|c_1| \ll |c_2| \sim |c_3| \sim |c_4|$ and
\begin{equation}
|c_1 \lambda_1^N| \sim |c_2 \lambda_2^N| \gg |c_3 \lambda_3^N| \gg |c_4 \lambda_4^N| \ \ , \ \ 
\label{orders_of_mag}
\end{equation}
where the sign $\sim$ represents the estimation within an order of magnitude. In more detail, the ratio
\begin{equation}
c_1\lambda_1^N/c_2\lambda_2^N \approx -[t_2]_4/[t_1]_4
\end{equation}
and the ratio $c_3/c_2$, $c_4/c_2$ are independent of $N$:
\begin{equation}
\begin{bmatrix}
     c_3/c_2 \\ c_4/c_2
   \end{bmatrix} \approx -\begin{bmatrix}
     [t_3]_2 & [t_4]_2 \\ [t_3]_4 & [t_4]_4
   \end{bmatrix}^{-1}  \begin{bmatrix}
     [t_2]_2 \\ [t_2]_4
   \end{bmatrix} \ \ . \ \ 
\end{equation}
From the separation of scale \eqref{orders_of_mag}, the complex impedance can be approximated as
\begin{equation}
Z^{(N)}_{AB} \approx  2Z_1\frac{[t_1]_1 [t_2]_4 - [t_1]_4 [t_2]_1}{[t_1]_2 [t_2]_4 - [t_1]_4 [t_2]_2} \bigg( 1 + \alpha \frac{\lambda_3^N}{\lambda_2^N} \bigg) \ \ , \ \
\label{Z_asymptotic}
\end{equation}
where $\alpha$ is given by:
\begin{equation}
\alpha =  \frac{c_3}{c_2} \left(\frac{[t_3]_1[t_1]_4 - [t_3]_4[t_1]_1}{[t_2]_1 [t_1]_4 - [t_2]_4 [t_1]_1} - \frac{[t_3]_2 [t_1]_4 - [t_3]_4[t_1]_2}{ [t_2]_2 [t_1]_4 - [t_2]_4 [t_1]_2 }\right)  \ \ . \ \
\label{Z_asymptotic}
\end{equation}
Given that $|\lambda_2|>|\lambda_3|$, $Z^{(N)}_{AB}$ converges to 
\begin{equation}
Z_{AB} = 2Z_1\frac{[t_1]_1 [t_2]_4 - [t_1]_4 [t_2]_1}{[t_1]_2 [t_2]_4 - [t_1]_4 [t_2]_2} \ \ . \ \ 
\label{convergence_value}
\end{equation}
Note that this value only depends on the eigenvectors $[t_1], [t_2]$ associated with the two eigenvalues that have the highest amplitudes. Physically speaking, the complex impedance of our infinite AC network comes from the two most dominant electrical modes propagating through the infinite series of unit cells.

\begin{figure*}[!htb]
\centering
\includegraphics[width=0.9\textwidth]{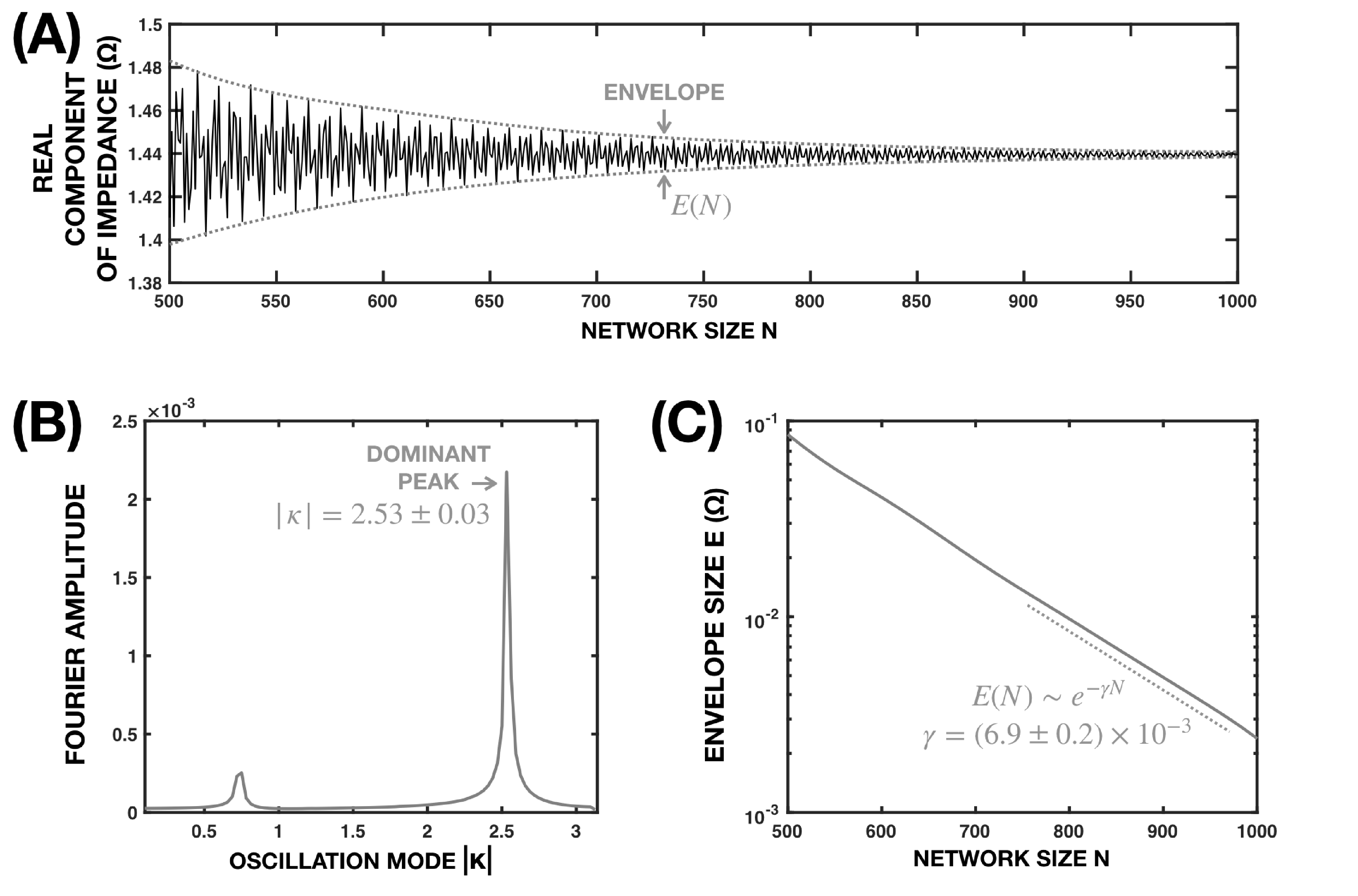}
\caption{(A) The real component of the complex impedance $\text{Re}\big( Z^{(N)}_{AB} \big)$ gradually converges to a fixed value. (B) Fourier transformation of $\text{Re}\big( Z^{(N)}_{AB} \big)$ in the range $N\in [800,1000]$. (C). The semilog plot of the envelope of $\text{Re}\big( Z^{(N)}_{AB} \big)$ in the range $N\in [800,1000]$.}
\label{convergence_damping_oscillator}
\end{figure*}

The equation \eqref{Z_asymptotic} shows that the convergence of $Z_{AB}^{(N)}$ has an exponential decay rate $\gamma$ and an oscillating frequency $|\kappa|$ which can be found from the identifications
\begin{equation}
\begin{split}
&e^{- \gamma N} \sim \big| \lambda_3^N/\lambda_2^N \big| \ \ \Rightarrow \ \ \gamma = \ln\big( |\lambda_2/\lambda_3| \big) \ \ , \ \ 
\\
&e^{i \kappa N} \sim e^{i\arg (\lambda_3^N/\lambda_2^N )} \ \ \Rightarrow \ \ |\kappa| = \big|\arg (\lambda_2/\lambda_3)\big| \ \ . \ \ 
\end{split}
\label{converge_features}
\end{equation}
Such behavior can be seen from numerical investigation as shown in Fig. \ref{convergence_damping_oscillator}A. The circuit investigated was the symmetric twisted ladder at frequency $\omega=2$rad/s, with inductors of $L= 1$H and capacitors of $C= 1$F, each connected in series with a resistance $R=0.01\Omega$. The highest non-zero peak in Fig.\ref{convergence_damping_oscillator}B appears to agree with $|\arg(\lambda_2/\lambda_3)| \approx 2.54$rad. The next highest peak is at $|\kappa| \approx |\arg(\lambda_2/\lambda_4)| \approx 0.74$rad. The rate of exponential decrease, or the slope of the best-fit line in Fig.\ref{convergence_damping_oscillator}C, is $\gamma = (6.9\pm 0.2)\times 10^{-3}$, which is not far off from $\ln(|\lambda_2/\lambda_3|) \approx 6.5\times 10^{-3}$.

When $|\lambda_2|=|\lambda_3|$, the convergence vanishes. When all four eigenvalues have the same magnitude, the situation becomes even worse because now there is no dominant electrical mode. Those two are the cases with our symmetric twisted ladder, $\omega < 1/\sqrt{LC}$ and $\omega \geq 1/\sqrt{LC}$ correspondingly, thus no convergence for all frequency $\omega$.

By adding an infinitesmal resistors $R=0^+\Omega$ in series with every AC elements:
\begin{equation}
\begin{split}
& Z_1 = i \omega L \rightarrow R + i \omega L \ \ , \ \ 
\\
&Z_2 = 1/i \omega C \rightarrow R + 1/i \omega C \ \ , \ \
\end{split}
\end{equation}
all eigenvalues will have different magnitudes. In more detail, for $\omega < 1/\sqrt{LC}$:
\begin{equation}
|\lambda_2| = 1 + |\epsilon| \ \ , \ \ |\lambda_3| = 1 - |\epsilon| \ \ , \ \ |\epsilon| \propto R \ \ , \ \  
\end{equation}
and for $\omega \geq 1/\sqrt{LC}$:
\begin{equation}
\begin{split}
& |\lambda_2| = 1 + |\epsilon| \ \ , \ \ |\lambda_3| = 1 - |\epsilon| \ \ , \ \ |\epsilon| \propto R  \ \ , \ \
\\
&|\lambda_1| = 1 + |\epsilon'| \ \ , \ \ 
|\lambda_4| = 1 - |\epsilon'| \ \ , \ \ |\epsilon'| \propto R \ \ , \ \ 
\end{split}
\end{equation}
where the sign $\propto$ represents a proportional relation. From \eqref{converge_features}, the rate of convergence $\gamma = 2|\epsilon| \propto R$, disappears when $R=0$ as expected. 

\subsection{On the Emergence of Real Impedance}

Using equation \eqref{convergence_value}, we arrive at the analytical result:
\begin{equation} 
\begin{split}
&Z_{AB}(\omega) = \lim_{R \rightarrow 0} \lim_{N\rightarrow \infty} Z^{(N)}_{AB}(\omega)= \sqrt{L/C}
\\
& \ \ \ \ \times \sqrt{ 1+2 \omega^2LC + 2i\omega \sqrt{LC} \sqrt{1-\omega^2LC}} \ \ , \ \
\end{split}
\label{theoretical_result}
\end{equation}
where the square-root value $\sqrt{\cdot}$ lies on the right half plane. This yields a good agreement in numerical investigation, as shown in Fig. \ref{numerical_investigation}B, under the condition:
\begin{equation}
N \gg \min (\omega L , 1/\omega C)/R \gg 1 \ \ . \ \ 
\label{limit_taking}
\end{equation}
When $\omega \geq 1/\sqrt{LC}$, the complex impedance has no imaginary component and thus becomes purely real:
\begin{equation}
\begin{split}
&Z_{AB}(\omega)\Big|_{\omega \geq 1/\sqrt{LC}} = \sqrt{L/C} 
\\
& \ \ \ \ \times \sqrt{1 +2 \omega^2LC - 2\omega \sqrt{LC} \sqrt{-1 + \omega^2LC}} \ \ . \ \ 
\end{split}
\end{equation} 
At a very large frequency $\omega \rightarrow \infty$, the complex impedence converges with a power law decay:
\begin{equation}
Z_{AB}(\omega) \approx \sqrt{2L/C} (1 + 1/16\omega^2LC) \rightarrow  \sqrt{2L/C} \ \ , \ \
\label{high_freq_convergence}
\end{equation} 
which can also be written as:
\begin{equation}
\lim_{\omega \rightarrow \infty} \lim_{R \rightarrow 0} \lim_{N\rightarrow \infty} Z^{(N)}_{AB}(\omega)= \sqrt{2L/C} \ \ . \ \ 
\end{equation} 
It is important to note that condition \eqref{limit_taking} indicates $N\rightarrow \infty$, $R \rightarrow 0$ and $\omega \rightarrow 0$ are noncommutative limits, i.e. another order of taking limits will give a different result. We should also mention that similar complications are also found in many different branches of physics, for example the Kolmogorov's four-fifths law in turbulence \cite{Frisch_Kolmogorov} with the averaging time $\mathcal{T}$, the kinematic viscosity $\nu$ and the stirring length scale $\mathcal{L}$:
\begin{equation}
\lim_{\mathcal{L}\rightarrow \infty} \lim_{\nu \rightarrow 0} \lim_{\mathcal{T}\rightarrow \infty} S_3(l) = -4 \epsilon l /5 \ \ , \ \ 
\end{equation}
where $S_3$ is the third order longitudinal structure function, $l$ is the distance between points of interests and $\epsilon$ is the energy dissipation per unit mass.

\section{Discussion}

This paper is our attempt to scratch the surface of a potentially deep topic. Not only can the complex impedance of an infinite AC network gain a real component by adding infinitesimal resistances, we now know that its imaginary component can be completely eliminated by entangling the topology. 

\begin{figure*}[!htb]
\centering
\includegraphics[width=0.6\textwidth]{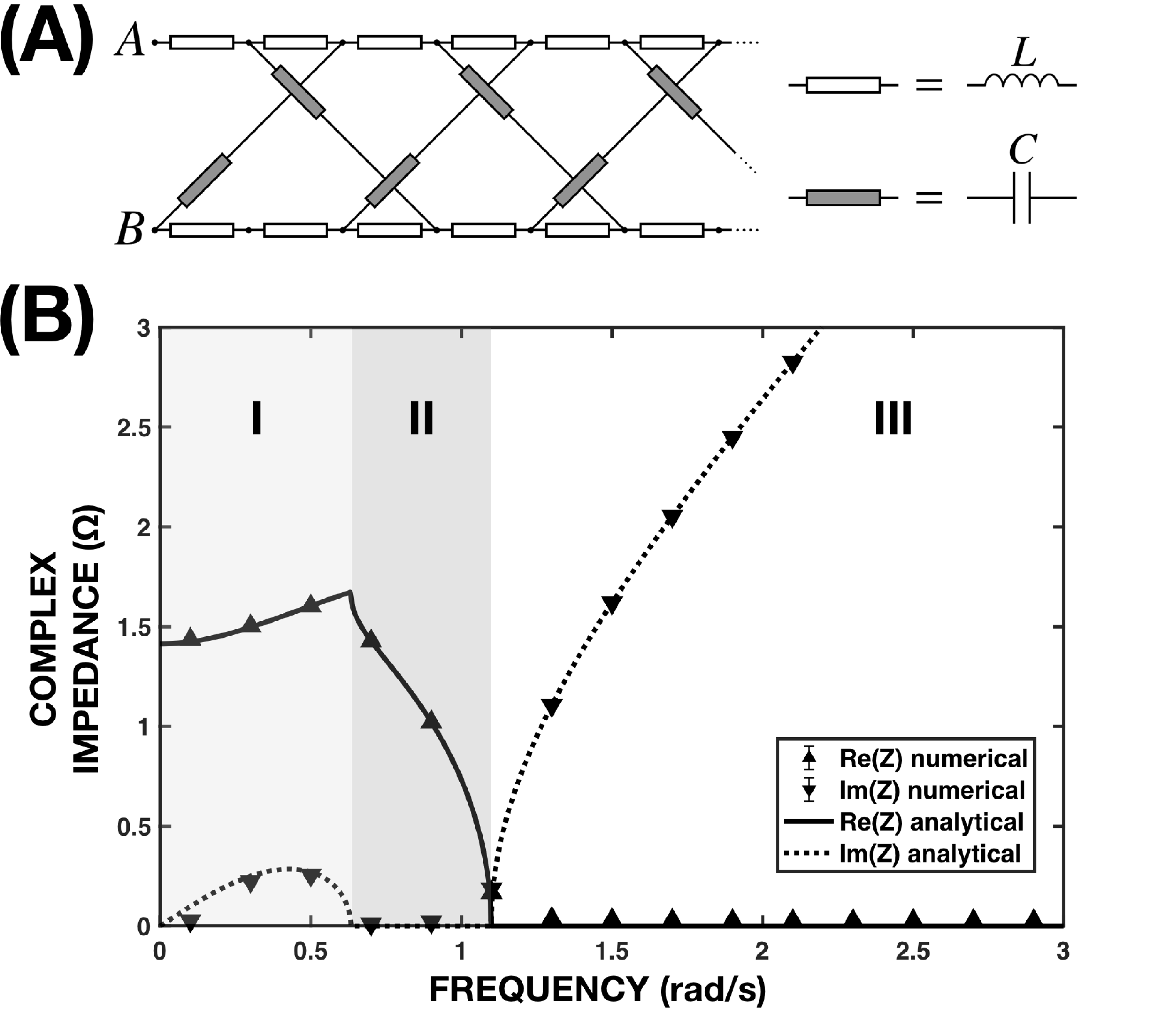}
\caption{(A) The asymmetric twisted ladder. (B) The average complex impedance in the range $N\in[450,500]$ versus the network frequency, when $L=1$H, $C=1$F, and each is connected in series to a resistance $R=0.01\Omega$. We also compare those with the analytical result for $N\rightarrow \infty$ and $R\rightarrow 0$ given in equation \eqref{theoretical_result_2}.}
\label{asym_twist_ac}
\end{figure*}

In our study, we also learn that one infinite AC network can exhibit all possible known behaviors with complex impedance. For example, consider another infinite AC network in Fig. \ref{asym_twist_ac}A, which we call the asymmetric twisted ladder:
\begin{equation}
\begin{split}
&Z_{AB}(\omega) = \lim_{R \rightarrow 0} \lim_{N\rightarrow \infty} Z^{(N)}_{AB}(\omega)= \sqrt{2L/C}
\\
& \ \ \ \ \times \sqrt{ 1+ \omega^2 LC + i\omega \sqrt{LC} \sqrt{2-5\omega^2LC}} \ \ . \ \
\end{split}
\label{theoretical_result_2}
\end{equation}
The result of our numerical investigation on this topology is given in Fig. \ref{asym_twist_ac}B. Here we can see all three distinct behaviors corresponding to regions I, II and III of the frequency $\omega$: (I) when $\omega < \omega_1$, $Z_{AB}(\omega)$ has both non-zero imaginary and real components; (II) when $\omega_1 \leq \omega \leq \omega_2 $ , $Z_{AB}(\omega)$ has only a real component; (III) when $\omega_2 \leq \omega$ , $Z_{AB}(\omega)$ has only an imaginary component. The frequencies $\omega_1$, $\omega_2$ are:
\begin{equation}
\omega_1 = \sqrt{2/5LC} \ \ , \ \ \omega_2 = \sqrt{(1 + \sqrt{2})/2LC} \ \ . \ \ 
\end{equation}

\begin{figure*}[!htb]
\centering
\includegraphics[width=0.8\textwidth]{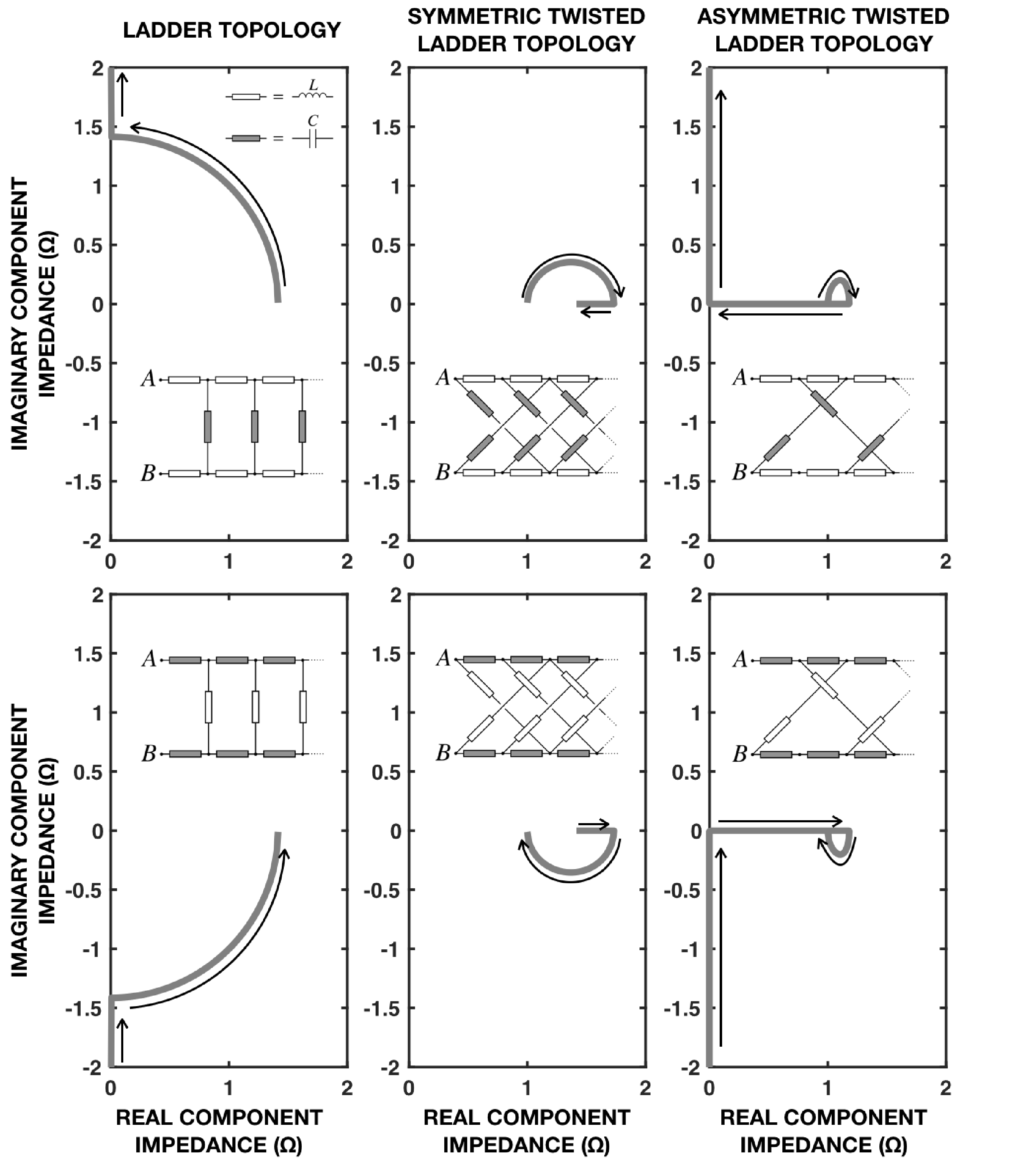}
\caption{The Nyquist plots of different infinite AC ladder networks, where the arrow indicates the evolution of $Z_{AB}(\omega)$ in the complex plane as the frequency goes from low to high $\omega = 0\rightarrow \infty$.}
\label{nyquist}
\end{figure*}

We summarize the ranges and trends of the complex impedances for some circuits in Fig. \ref{nyquist}, using Nyquist plots -- a method often used in electrochemical impedance spectroscopy \cite{Macdonald_Electrochem}. We have not observed any infinite real impedance but only observed infinite imaginary parts. 

Here, we also link some features of these plots to some features in the topologies of the corresponding circuits based on our limited data, and attempt some qualitative explanations which we find useful intuitively. The complex impedance remains finite in the top-center, bottom-left and bottom-right circuits in Fig. \ref{nyquist} even when the frequencies approach infinity. We notice that these three circuits differ from the rest in the presence of ``capacitor chains'': there are many ways ($\sim N$) to connect the two terminals by capacitors with only a few inductors in between. When the frequency becomes large, the impedance of the capacitors vanishes, and the impedances of those chains are of the order of $j\omega L + N/j\omega C + NR$, orders of magnitude smaller than the $jN\omega L$ of ``inductor chains''.

Hence, using the random-walk analogy of electrical circuits \cite{Doyle_Random_Walk}, the two circuits can be viewed as containing $\sim N$ paths with large probabilities between the two terminals compared to other circuits. Therefore, the overall probability connecting to the circuit should also be larger. A very rough estimate would add all the probabilities of such paths together. After converting escape probability to impedance, it gives $j\omega L/N + R$. We clearly overdid it, but at least this estimation shows a hint of why circuits of these types are expected to have finite complex impedances in the limit of interests. Note that the original analogy is for real impedances only, but here we extend it using naive analytical continuation.

There are many unanswered questions and features of infinite AC networks that we still do not fully understand: the ties to graph theory and how order of magnitude estimates of the limiting behaviors of the impedance can be made just based on the features of the circuit, how the network's topology decides (conclusively) which behavior of $Z_{AB}$ will be shown at what range of frequency $\omega$, how the convergence rate $\gamma$ depends on the network's topology, can the choice of adding infinitesimal resistance to every AC elements always lead to $Z^{(N)}_{AB}$ convergence, can the choice of posing boundary conditions at the infinitely far end change $Z_{AB}$ in a drastic manner, can a negative real component of complex impedance $\text{Re}(Z_{AB}) < 0$ ever emerge ... The list goes on, and we will have to leave those curiosities to future research.

\begin{acknowledgments}

We thank Hiep T. Vu, Duy V. Nguyen, and xPhO club for their valuable feedback and insightful comment throughout every stage of this research.

\end{acknowledgments}

\bibliography{MainNotes}
\bibliographystyle{apsrev4-2}

\end{document}